# On the culture of open access: the Sci-hub paradox

Abdelghani Maddi[1] and David Sapinho[2]

[1] abdelghani.maddi@hceres.fr
Observatoire des Sciences et Techniques, Hcéres, 2 Rue Albert Einstein, Paris, 75013 France ;
GEMASS, CNRS - Sorbonne University, 59/61 rue Pouchet, 75017, Paris, France

[2] david.sapinho@hceres.fr
Observatoire des Sciences et Techniques, Hcéres, 2 Rue Albert Einstein, Paris, 75013 France

## Abstract

Shadow libraries, also known as "pirate libraries", are online collections of copyrighted publications that have been made available for free without the permission of the copyright holders. They have gradually become key players of scientific knowledge dissemination, despite their illegality in most countries of the world. Many publishers and scientist-editors decry such libraries for their copyright infringement and loss of publication usage information, while some scholars and institutions support them, sometimes in a roundabout way, for their role in reducing inequalities of access to knowledge, particularly in low-income countries. Although there is a wealth of literature on shadow libraries, none of this have focused on its potential role in knowledge dissemination, through the open access movement. Here we analyze how shadow libraries can affect researchers' citation practices, highlighting some counter-intuitive findings about their impact on the Open Access Citation Advantage (OACA). Based on a large randomized sample, this study first shows that OA publications, including those in fully OA journals, receive more citations than their subscription-based counterparts do. However, the OACA has slightly decreased over the seven last years. The introduction of a distinction between those accessible or not via the Sci-hub platform among subscription-based suggest that the generalization of its use cancels the positive effect of OA publishing. The results show that publications in fully OA journals are victims of the success of Sci-hub. Thus, paradoxically, although Sci-hub may seem to facilitate access to scientific knowledge, it negatively affects the OA movement as a whole, by reducing the comparative advantage of OA publications in terms of visibility for researchers. The democratization of the use of Sci-hub may therefore lead to a vicious cycle, hindering efforts to develop full OA strategies without proposing a credible and sustainable alternative model for the dissemination of scientific knowledge.

## Acknowledgement

The present paper is a substantially extended version of the contribution (Maddi and Sapinho, 2022b) presented at the 26th International Conference on Science, Technology and Innovation Indicators (STI 2022), Granada, Spain, 7-9 September 2022. A pre-print version of this manuscript is available on the Research Square platform, available at: https://www.researchsquare.com/article/rs-2357492/v1. The authors would like to thank the reviewers for their insightful comments and suggestions that have contributed to the improvement of the quality of the present paper.

## Compliance with Ethical Standards

The authors have no relevant financial or non-financial interests to disclose.

# Introduction

Also known as "pirate libraries", shadow libraries are online collections of copyrighted publications that have been made available for free without the permission of the copyright holders. Over the past decade, shadow libraries have undergone a dramatic evolution, both in terms of coverage of indexed publications and use by the scientific community. Several factors have contributed to their progression, led by the launch of the largest shadow library Sci-hub, in 2011 (Bohannon, 2016). The notoriety of this pirate site has grown so fast that in 2016 Alexandra Elbakyan – founder of Sci-Hub – was ranked in the top 10 influential figures of Nature (Nature's 10, 2016, p. 10). The great ease of use and the indexing of the majority of publications, including those under copyright, make Sci-hub a tool used more and more regularly by researchers.

Although this type of libraries (e.g. Library Genesis, Z-Library or Sci-Hub) share the same objective as that of open access advocates (make science accessible to all without limitation), the means are not the same (Banks, 2016). This is not without consequences. Notwithstanding the loss of information on the use of scientific publications (McNutt, 2016), some consider that the massive use of these libraries risks collapsing the current publishing market (Himmelstein *et al.*, 2018).

Since its launch, shadow libraries, particularly Sci-hub, have been the subject of much discussion in the literature, under several prisms. In addition to descriptive studies on its content and its use by researchers (Greshake, 2017; Himmelstein *et al.*, 2018; Karaganis, 2018; Nicholas *et al.*, 2019; Till *et al.*, 2019), some contributions have focused on its role in reducing inequalities of access to scientific knowledge (Boudry *et al.*, 2019; Bodó, Antal and Puha, 2020), or even on its impact on the citations received by publications (Correa *et al.*, 2022). However, we are not aware of any study on the impact of Sci-hub on the Open Access (OA) movement itself. By breaking down barriers to access to scientific knowledge, does Sci-hub promote the OA movement, or on the contrary does it slow it down? In other words, by providing access to subscription-based publications illegally, Sci-hub could cancels the virtues of OA publishing (legal way) and risks acting against the OA movement. From a bibliometric point of view, the massive use of shadow libraries like Sci-hub risks canceling the Open Access Citation Advantage (OACA) in favor of non-OA journals, which are generally older and more established in the scientific publishing market.

In this paper, we analyze how shadow libraries, more precisely Sci-hub, can affect researchers' citation practices, and the impact this can have on the OACA both for publications in fully OA journals and in hybrid journals.

In the previous version of this paper (Maddi and Sapinho, 2022b), we showed that the OACA does exist for OA publications published in hybrid journals, unlike publications in fully open access journals, which rather present a citation disadvantage. We hypothesized in the directions for further research that Sci-hub may have a role in researchers' citation practices. Therefore, the aim of this paper is to analyze how Sci-hub can affect this advantage/disadvantage.

# Literature review on the OACA

OACA has been a major topic of discussion in the literature over the past twenty years, generally assuming that a better accessibility fosters research impact. In fact, this belief is not so obvious, with contradictory findings in the studies dealing with the OACA, that have often proven tough to compare, depending on many factors such as disciplines, OA status taken into account, publications types as well as the method and database used.

The first paper to analyze the question of OACA is that of Steve Lawrence (Lawrence, 2001), published at the turn of the century in Nature. He made a correlation between the number of citations and the share of freely accessible articles. He found a positive correlation between the two indicators. Since then, dozens of papers have been published on the topic. Overall, studies that found the existence of OACA were more common in social sciences (Mikki *et al.*, 2018; Abbasi *et al.*, 2019; Valderrama-Zurián, Aguilar-Moya and Gorraiz, 2019), Medical and health sciences (Hudson *et al.*, 2019; O'Kelly, Fernandez and Koyle, 2019; Miller *et al.*, 2021), and Natural sciences (Lin, 2007; Wang *et al.*, 2015; Clements, 2017). OACA is less important (and in several disciplines non-existent) in Physical Sciences and Engineering (Archambault *et al.*, 2016). In addition, there are some studies that concluded on the nonexistence of OACA in social sciences (Zhang, 2006), Medical and health sciences (Tonia *et al.*, 2016; Mimouni *et al.*, 2017; Nazim and Ashar, 2018), and Natural sciences (Campos *et al.*, 2016; Peidu, 2020). On a large sample of the Web of Science database, Dorta-Gonzalez et al. (2017)(Dorta-González, González-Betancor and Dorta-González, 2017) concluded that there is no OACA in all disciplines.

Recently, a review analysis on the topic identified 134 publications dealing with OACA (Langham-Putrow, Bakker and Riegelman, 2021). Applying the EBL critical appraisal method to analyze the risks of bias based on factors like sample size, data collection or study design, the authors emphasized that most of these studies (131 – about 98%) present a high risk of bias. Two of the three publications with a low risk relate to medical and natural science research (Tonia *et al.*, 2016; Nelson and Eggett, 2017), and the third (Sotudeh, Ghasempour and Yaghtin, 2015) used relatively old data (2007-2011). Moreover, none of these articles used randomization techniques/control group.

Given that several confounding factors have limited the scope of results from existing studies, it was not possible to conclusively confirm the existence or non-existence of the OACA. A key issue to address the question rigorously would be to figure out a way to "isolate" the OA effect. OA publications should thus be compared to a counterfactual sample of publications with the only difference is to be published in subscription-based journals.

## Method

### Data

We extracted the publications data from the French OST in-house database. It includes five indexes of the WoS available from Clarivate Analytics (Science Citation Index Expanded (SCIE), Social Sciences Citation Index (SSCI), Arts & Humanities Citation Index (AHCI), Conference Proceedings Citation Index (CPCI-SSH) and Conference Proceedings Citation Index (CPCI-S)), and corresponds to WoS content indexed through the end of March 2021. The study focuses on three types of publications: articles, reviews and conference proceedings.

### Sample selection procedure and weighting

When working with samples (e.g. public opinion polls), it is important to adjust data for population parameters such as gender, age, region, etc. We propose in this work to adjust structure of the two non-OA control groups by calculating weights so that these samples corresponds to that of OA publications (respectively in fully OA and hybrid journals). In other words, we calculated the raking weights for the publications of the control sample in order to ensure that both samples have a comparable distribution with respect to the characteristics identified. In our calculations, we distinguish publications in fully OA journals and those in hybrid journals.

There are several ways to do this. One of them is the Raking Ratio method. We used Josh Pasek's package (see: https://cran.r-project.org/web/packages/anesrake/index.html) to calculate raking weights in R.

### OA sample

We selected all the documents published with a Gold OA status from 2010 to 2020 by distinguishing those published in a fully OA and those in hybrid journals, representing respectively 2,458,378 and 1,024,430 publications.

We limit ourselves to publications in gold OA, but distinguishing between publications in fully OA journals and hybrid journals. The reason why we did not take into account the green and bronze routes is the difficulty of conducting a large-scale analysis of the OACA for this type of publications. Thus, for the green route, it is not possible to identify precisely the moment when each publication was dumped in an open archive, when for the bronze route, the durability of the OA status is highly questionable. For each non-OA publication, we identified (using DOI) whether its full text is available in Sci-hub (which indexes in December 2021, 88,343,822 publications). Thus, we calculated the weighted Mean Normalized Citation Score (MNCS) (Leydesdorff and Opthof, 2010) for the two control groups depending on whether the publications are in Sci-hub or not. Finally, we calculated the OACA for publications in fully OA journals and for those in hybrid journals (taking into account whether non-OA are in Sci-hub or not).

### Control samples

We used the raking ratio method (Deville and Särndal, 1992; Deville, Sarndal and Sautory, 1993; Sebastian Daza, 2012) to ensure comparability between the two samples. The method comprises the construction of a control sample similar to the sample of interest, except for the analyzed parameter, which is in our case the OA status.

To do so, we first propose a method to constitute a control group to isolate the OA effect. By definition, OACA analysis answers the following question: "what would be the impact of an OA publication if it had been published in a subscription-based journal?" To answer this question it is necessary to compare this publication to its counterparts that share the same characteristics that can affects citation impact, such as the discipline, journal visibility and other known characteristics (e.g. the number of authors, whether it has received a specific funding or the number of funders). We thus constructed two non-OA control groups for publications in fully OA journals and for publications in hybrid journals. Once our control groups were constructed, we applied sample adjustment methods (raking ratio) so that their structure matched that of the OA publications.

The control samples are thus obtained by finding all the non-OA publications, qualified as doubles, that match with OA publications (depending on whether they are published in fully OA journals or hybrid journals) on a set of features identified in the literature as having an effect on the citation impact (Judge *et al.*, 2007; Yan, Wu and Song, 2018; Waltman and Traag, 2021; Maddi and Sapinho, 2022a). The main publications characteristics used for raking ratio (see method) are:

- the publication year (11 classes : 2010 to 2020),
- the discipline (OST classification in 27 ERC panels),
- the journal impact (5 classes : <0.8, [0.8 , 1.2[, [1.2 , 1.8[, [1.8 , 2.2[, >=2.2), for the calculation method see: (Maddi and Sapinho, 2022a).
- the number of countries of contributors, based on WoS addresses information (5 classes : 1,2,3,4 and 5 or more),
- the number of funding received, based on WoS acknowledgment information (5 classes : 1,2,3,4 and 5 or more),
- the presence of an European Research Council (ERC) funding (2 classes : Yes or No),
- the presence of at least one European (UE27) address (2 classes : Yes or No),
- the presence of a patent citation (2 classes : Yes or No).

On this basis, we categorized each OA publication in one among 242,924 different clusters. From the same clusters, we then identify 12,088,681 double candidates among which 10,310,342 and 11,533,001 are respectively eligible for full OA and hybrid OA publications.

## OACA calculation

We define the Open Access Citation Advantage (OACA) as the percentage of advantage in the citation score that could be attributed to OA. We used the following formula:

$$OACA = \frac{MNCS_{OA} - MNCS_{(Non-OA)_{ij}}}{MNCS_{(Non-OA)_{ij}}} * 100$$

Where $MNCS_{OA}$ is the Mean Normalized Citation Score (Leydesdorff and Opthof, 2010) in the OA group, and $MNCS_{Non-OA}$, the same score in the Non-OA control group. "i" represents the type of the control group (for publications in fully OA or in hybrid journals), "j" represents the presence or not of the full text in Sci-hub. Thus, we calculate four OACAs summarized in the table 1.

**Table 1: OACA by type of OA journal and according to the presence of non-OA in Sci-hub**

| #OACA | Journal type of OA publications | Non-OA control group type | Non-OA is available in Sci-hub |
|---|---|---|---|
| 1 | Fully OA | Control group for publications in fully OA journals | Yes |
| 2 | | | No |
| 3 | Hybrid | Control group for publications in hybrid journals | Yes |
| 4 | | | No |

Figure 1 summarizes the sample selection procedure and the OACA calculation.

**Figure 1: Method of constructing the control samples of OA publications, and OACA calculation**

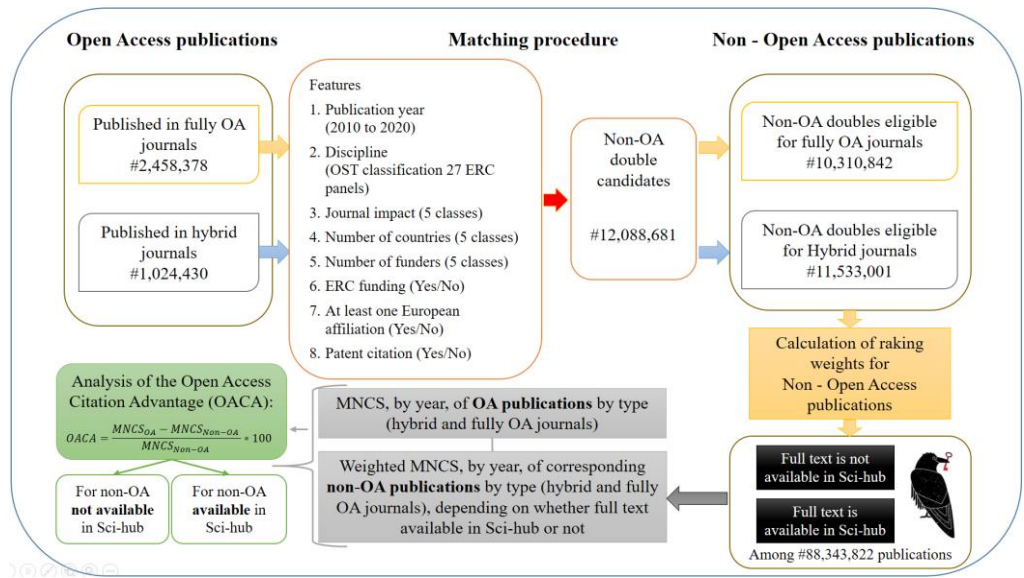

## Results

In this section, we first present some descriptive statistics on Sci-hub coverage, then the main results on the OACA by type.

### Descriptive statistics

Figure 2 shows the evolution of the share of non-OA publications whose full text is available in Sci-hub. Of the 12,088,681 non-OA publications, Sci-hub indexes 63%. Over the 2009-2020 period, the share of non-OA publications downloadable from Sci-hub increased from about 55% to more than 75%.

**Figure 2: Share of non-OA publications available on Sci-hub, by year (2009-2020)**

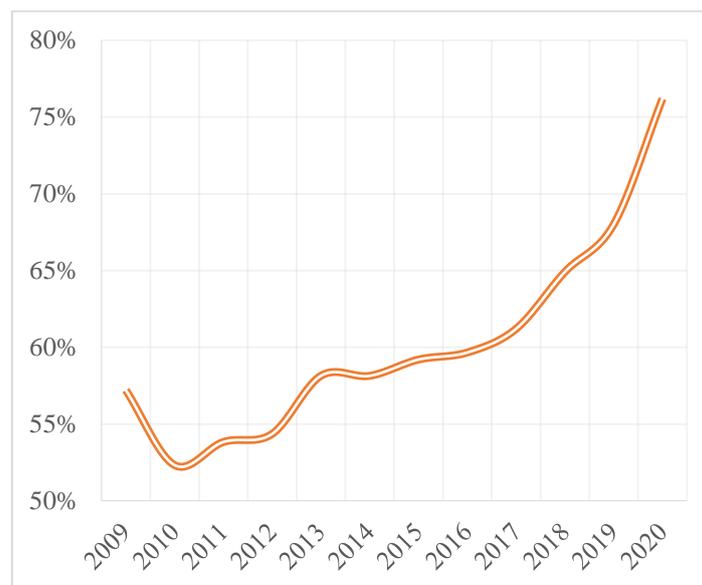

Figure 3 shows the share of non-OA publications available on Sci-hub, by discipline. There are many disparities in the indexing of publications by discipline. If the coverage of Sci-hub is very high in Life Sciences, it is much less so in Social Sciences and Humanities and in Physical

Sciences and Engineering. On the latter, PE4 (Physical and Analytical Chemical Sciences) and PE5 (Synthetic Chemistry and Materials) are exceptions with a very high coverage; i.e. shares of 83.6% and 79.7% respectively. Sci-hub provides access to almost all non-OA publications in LS3 (Cellular, Developmental and Regenerative Biology), LS2 (Integrative Biology: From Genes and Genomes to Systems) and LS5 (Neuroscience and Disorders of the Nervous System). Conversely, the share of publications whose full text is available in Sci-hub is very low in PE7 (Systems and Communication Engineering) and PE6 (Computer Science and Informatics) with only 17.4% and 31.7% respectively. This is to be expected because these disciplines publish many conference proceedings, many of which have no DOI. Social Sciences and Humanities coverage in Sci-hub is also relatively low, particularly in SH5 (Cultures and Cultural Production) and SH6 (Study of the Human Past) with shares of 41.2% and 46.4% respectively.

**Figure 3: share of non-OA publications available on Sci-hub, by discipline, 2018-2020**

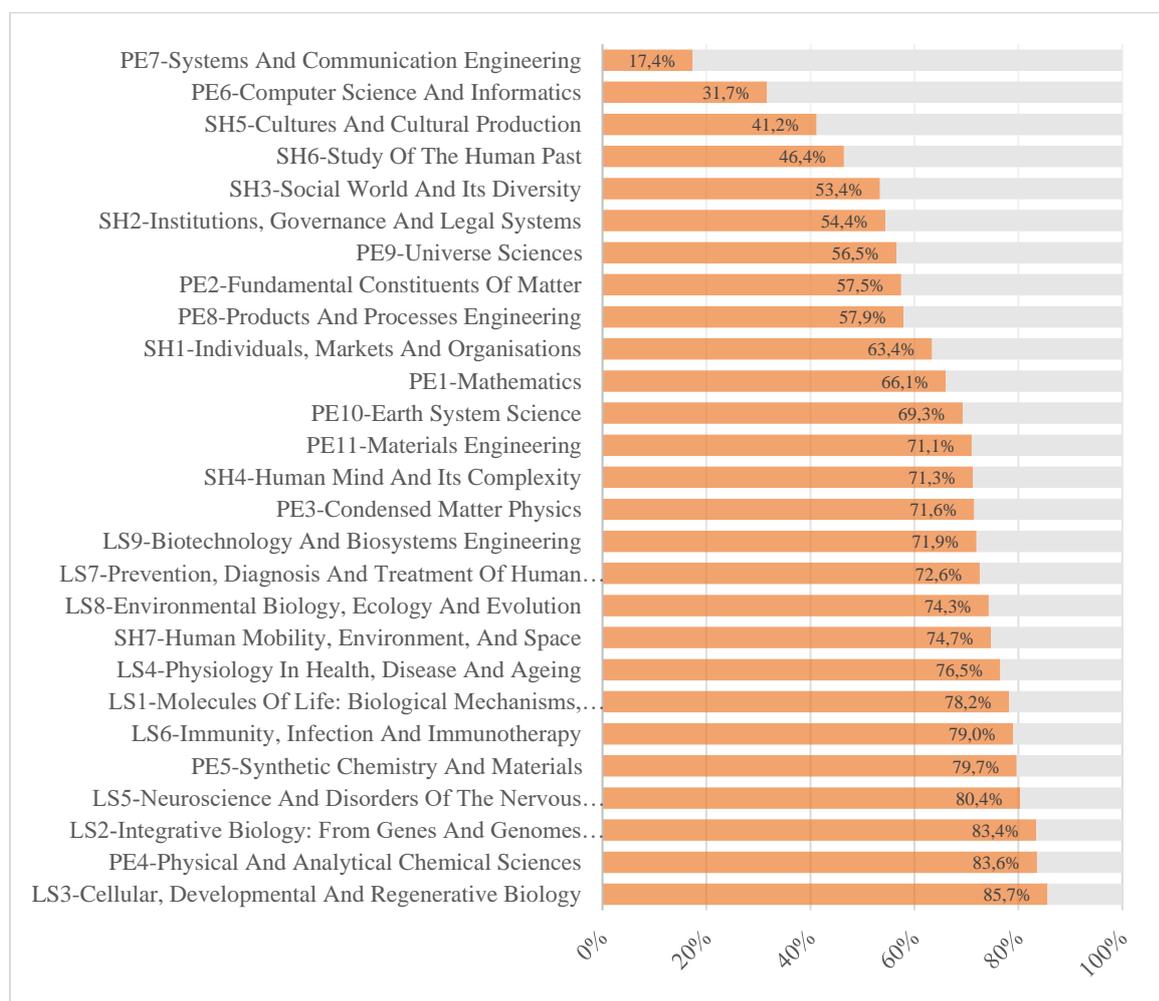

### The OACA analysis

In this section, we show the results on the effect of Sci-hub on the OACA. We can draw several lessons from Figure 4.a. First, we find that for publications in fully OA journals, there is a citation advantage over publications in non-OA ones, but on condition that the latter are not available in Sci-hub. This would mean that access to the full text of publications is an essential element in citation practices, regardless of the type of journal (fully OA or non-OA). This is especially true, as the OACA becomes a disadvantage if non-OA publications are available in Sci-hub. A second interesting observation is the gradual reduction of the citation disadvantage between 2009 and 2016. We could explain that by the fact that OA journals have gained notoriety over time and are attracting high-impact senior researchers. Finally, there is a drop in the OACA from 2016 that we cannot explain with certainty. Nevertheless, we assume that it is related to an increase in the number of newly created OA journals (or newly indexed in the WoS database) from 2016. Another possible hypothesis is that the proliferation of institutional and subject repositories during this period (after 2016) may have played a role. Non-open access articles may now be accessible through green open access road, which could potentially reduce the citation advantage of gold open access. This may be subject to future development.

**Figure 4: Evolution of the OACA by type of OA journal and according to the presence of non-OA publications in Sci-hub**

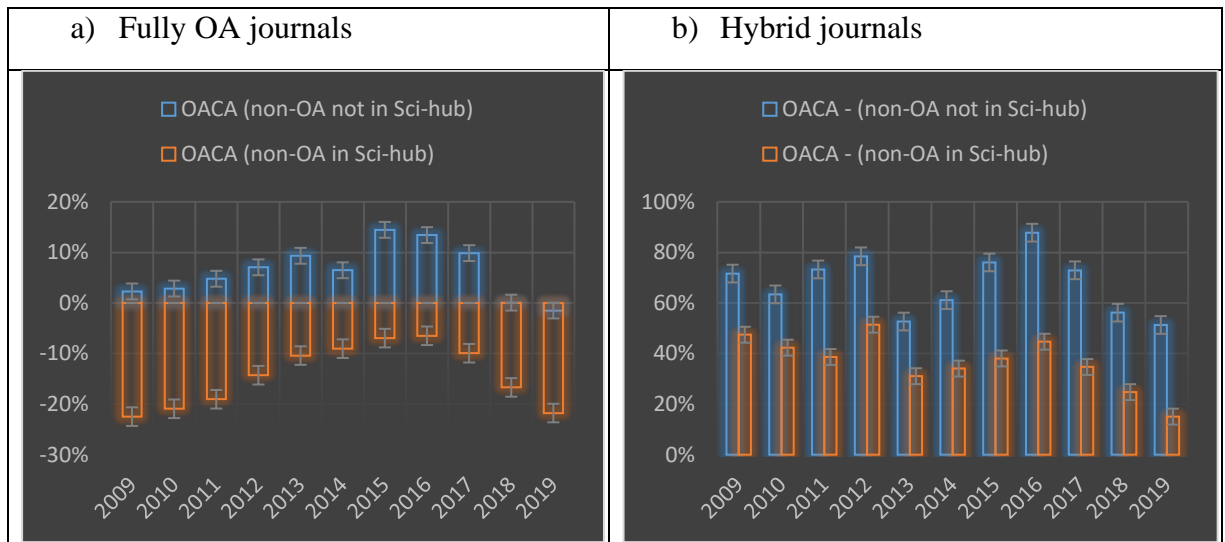

In figure 4.b, we observe the same trend; OACA is significantly higher if non-OA publications are not available in Sci-hub. What is interesting to point out here is that for OA publications in hybrid journals, the OACA remains positive in both cases, even if we also observe the drop from 2016. In other words, OA publications in hybrid journals tend to be cited more than non-OA publications whether the latter are in Sci-hub or not. Two types of "bias" may be behind this result: the first one is the selection bias: hybrid journals are often prestigious and well-known journals, and authors may submit their best work to these journals. In contrast, open access journals may be perceived as less selective, which could reduce the quality of the published articles. the second one is the reputation bias: hybrid journals often have a long history and established reputation in their field. Researchers may be more inclined to cite articles published in these journals rather than in newer or less well-known open access journals.

## Discussion

### On the OACA results

The lack of OACA for publications in fully OA journals is to be expected, as a great proportion of OA journals are newly created and less attractive to high-impact senior researchers. However, the citation disadvantage of these publications has largely diminished over time to disappear in 2016, as these journals gain a readership over the years and become more visible. OACA of publications in hybrid journals is also to be expected. These journals are well established in the publishing market and highly visible to scientific communities. For this type of journal, the fact that an article is in OA has a positive impact on citations more than articles that are in the same journals but with limited access to subscribers (hence the OACA). These results therefore suggest that the OACA depends on the type of journal; i.e. to hope to benefit from an OACA, it would first be necessary to choose a journal visible to the community (high-impact journal). In this case, the OA status will act as a catalyst that would increase the citations received.

### On the Sci-hub effect

One of the most striking results of this study is the role of shadow libraries (Sci-hub) in the citation practices of researchers. Although their goal is to promote open access, these types of libraries unwittingly work against the OA movement. The results show that publications in fully OA journals (and to a lesser extent those in hybrid journals) are victims of the success of Sci-hub. The latter overshadows them and prevents their development, insofar as the issue of easy access to scientific knowledge on which they are built is blurred with the development of dark OA. This paradoxical situation that characterizes Sci-hub (and shadow libraries) deserves a fundamental debate even among free access advocates to have a clear position on this type of library and enlighten the scientific community on the consequences of their massive use.

## Conclusion

Through this paper, we sought to analyze the role of Sci-hub on authors' citation practices and consequently on the OACA (Open Access Citation Advantage). In other words, does the massive use of this pirate site impact the citation practices of researchers by increasing the visibility of non-OA publications to the detriment of OA ones?

To do so, we compared citation impact (MNCS) of 2,458,378 publications in fully OA journals to that (weighted MNCS) of a control group of non-OA publications (#10,310,842). Similarly, we did the same exercise for OA publications in hybrid journals (#1,024,430) and their control group (#11,533,001), over the period 2010-2019. Each time we have distinguished the non-OA publications, which are available in Sci-hub, and those that are not.

The results show that Sci-hub indexes the majority of scientific publications as the share of non-OA publications available in Sci-hub increased from 55% to 76% between 2009 and 2019. This progression negatively affects the OACA. Thus, publications in fully OA journals receive on average more citations than their equivalents accessible by subscription (non-OA), on the condition that the latter are not available in Sci-hub. We observed the same trend for publications in hybrid journals; OA publications in these journals receive on average more citations than non-OA ones. As for publications in OA journals, in hybrid journals the OACA is less if subscription-based articles are available in Sci-hub. Therefore, this study showed for the first time, on a large-scale randomized analysis, that the OACA does exist, and that the development of the dark route of open access tends to reduce it.